\documentclass[aps,prl,showpacs,preprint]{revtex4}

\usepackage{amsmath}

\usepackage{graphicx}
\usepackage{dcolumn}
\usepackage{bm}
 
\DeclareMathOperator\erfi{erfi}
\DeclareMathOperator\erf{erf}

\begin{document}

\title{Kinetic Equation and Non-equilibrium Entropy for a Quasi-two-dimensional Gas }
\author{J.Javier Brey, Pablo  Maynar, and M.I. Garc\'{\i}a de Soria}
\affiliation{F\'{\i}sica Te\'{o}rica, Universidad de Sevilla,
Apartado de Correos 1065, E-41080, Sevilla, Spain}
\date{\today }

\begin{abstract}
A kinetic equation for a dilute gas of hard spheres confined between two parallel plates separated a distance smaller than two particle dimeters is derived. It is a Boltzmann-like equation, which incorporates the effect of the confinement on the particle collisions. A function $S(t)$ is constructed by adding to the Boltzmann expression a confinement  contribution. Then it is shown that for the solutions of the kinetic equation, $S(t)$ increases monotonically in time, until the system reaches a stationary inhomogeneous state, when $S$ becomes the equilibrium entropy of the confined system as derived from equilibrium statistical mechanics. From the entropy, other equilibrium properties are obtained, and Molecular Dynamics simulations are used to verify some of the theoretical predictions.

\end{abstract}


\maketitle
Fluids confined between two parallel plates are intermediate between fluids in two and three dimensions. The confinement can strongly influence the physical properties of the system as compared with the three-dimensional bulk behavior and also with the two-dimensional limiting case.  In the last decades, the study of confined fluids have attracted a lot of attention,  mostly focussed on equilibrium and phase transitions properties
\cite{Th92,DyH95,SyL96,FLyS12,TMyE87,RSLyT97}. On the other hand, although it is true that in some cases the  dynamical behavior has been addressed \cite{KyD89,Letal10}, few non-equilibrium results seem to be well established  in the context of a general theory.  Recently, the interest in the slit geometry with two parallel, hard plates of a single component 
system of hard spheres, with a separation between the two plates smaller than two particle diameters, has largely increased, due to a series of experimental observations in systems of macroscopic particles \cite{Metal05,RIyS06,Retal11,CMyS12}. Although it is quite sure that the inelasticity of collisions, inherent to the macroscopic character of the spheres, and the subsequently needed  energy injection to reach and maintain a steady state, play a crucial role in many of the observed features \cite{BRyS13,BGMyB13,BBMyG15}, it is clear that before considering the consequences of these factors, the idealized system of elastic hard spheres must be addressed. The central issue is whether it is possible to formulate a macroscopic, hydrodynamic-like theory to describe the two-dimensional dynamics of the system when observed from above or from below and, if the answer is affirmative, which is the form of the equations and the expressions of the coefficients appearing in them.  Kinetic theory and non-equilibrium statistical mechanics provide the adequate framework to address these issues.

\begin{figure}
\includegraphics[width=.7\textwidth]{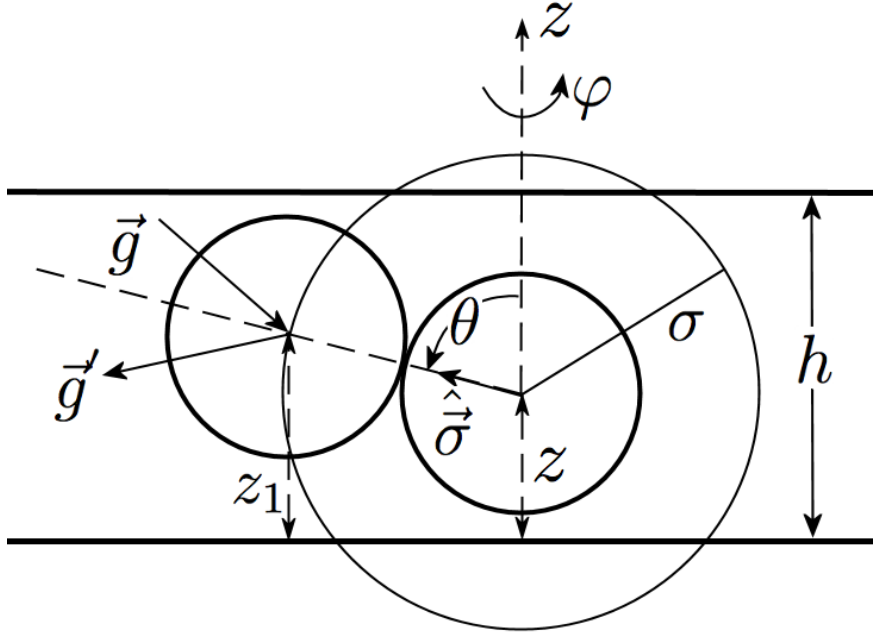}
  \caption{Collision between two hard spheres in a quasi-bidimensional confined system. The collision vector $\widehat{\bm \sigma}$ is univocally determined  by the coordinates $z$, $z_{1}$, and the azimuth angle $\varphi$.}
  \label{fig1}
\end{figure}

The present work deals with the formulation of a kinetic equation for a quasi-two-dimensional system of $N$ hard spheres of diameter $\sigma$ in the low density limit, with the same degree of validity and accuracy as the Boltzmann-Enskog equation, that lies at the heart of the theory of the non-equilibrium behavior  in many relevant applications of statistical mechanics \cite{Ce88,Kr10}. Its extension to higher densities starting from the pseudo-Liouville equation and the formulation of an empirical kinetic equation in the spirit of the Enskog approximation \cite{EDHyvL69} is straightforward. The system is confined between two large, formally infinite parallel plates located at $z=0$ and $z=h$, $\sigma < h < 2 \sigma$. This particular geometry allows to fix in a biunivocal way the position of two particles at contact by means of their $z$ coordinates and the polar angle $\varphi$ around the $z$ axis, as sketched in the Fig. \ref{fig1}. It is assumed that in the low density limit, the one-particle distribution function $f({\bm r},{\bm v},t)$ of the gas can be considered as constant over displacements of the order of the diameter $\sigma$ in any direction parallel to the walls, i.e. in the $x-y$ plane. On the other hand, the same can not be expected to hold in the $z$-direction, since the confinement imposed by the plates occurs over a distance smaller  than $\sigma$ and the isotropy of the dynamics of collisions is broken. Moreover, it must be emphasized that the relevant dimensionless parameter defining the low  density limit in which the equation applies is not the three-dimensional one but the (effective) two-dimensional density. Then, using the Stosszahlansatz or molecular chaos hypothesis, i.e. assuming that there are no correlations between the dynamical states of particles before  collisions, and the usual arguments leading to the Boltzmann-Enskog equation \cite{McL89}, one gets
\begin{equation}
\label{1}
\frac{\partial f}{\partial t} +{\bm v} \cdot \frac{\partial f}{\partial {\bm r}}= J[{\bm r},{\bm v} |f],
\end{equation} 
where the collision term $J$ is given by
\begin{eqnarray}
\label{2}
J[{\bm r},{\bm v}|f] &\equiv&  \sigma^{2} \int d{\bm v}_{1}  \int_{\sigma /2}^{h-\sigma/2} dz_{1} \int_{0}^{2\pi} d \varphi\, 
|{\bm g} \cdot \widehat{\bm \sigma}| \left[  \theta ({\bm g} \cdot \widehat{\bm \sigma} ) f(x,y,z_{1},{\bm v}_{1}^{\prime},t)f({\bm r},{\bm v}^{\prime},t) \right. \nonumber \\ 
&& \left. -  \theta (- {\bm g} \cdot \widehat{\bm \sigma} ) f(x,y,z_{1},{\bm v}_{1},t)f({\bm r},{\bm v},t) \right].
\end{eqnarray}
Here, ${\bm g} \equiv {\bm v}_{1} -{\bm v}$, ${\bm r} \equiv \{ x,y,z \}$, $\theta$ is the Heaviside step function, $\widehat{\bm \sigma} (z-z_{1}, \varphi)$ is the unit vector along the line joining the centers of the two particles at contact, and ${\bm v}^{\prime}$ and ${\bm v}_{1}^{\prime}$ are the velocities of the two hard spheres after the collision. 
Upon deriving the above equation, it has been assumed that no external force is acting on the system. Moreover, the walls confining the system are considered to be hard and at fixed positions. In this case, their effect can be described by means of boundary conditions to the kinetic equation \cite{DyvB77}. Let us mention that, in principle,  Eq. (\ref{1}) can be generalized for an arbitrary separation of the two parallel plates, although the mathematical description becomes much more involved.

\begin{figure}
\includegraphics[width=.7\textwidth]{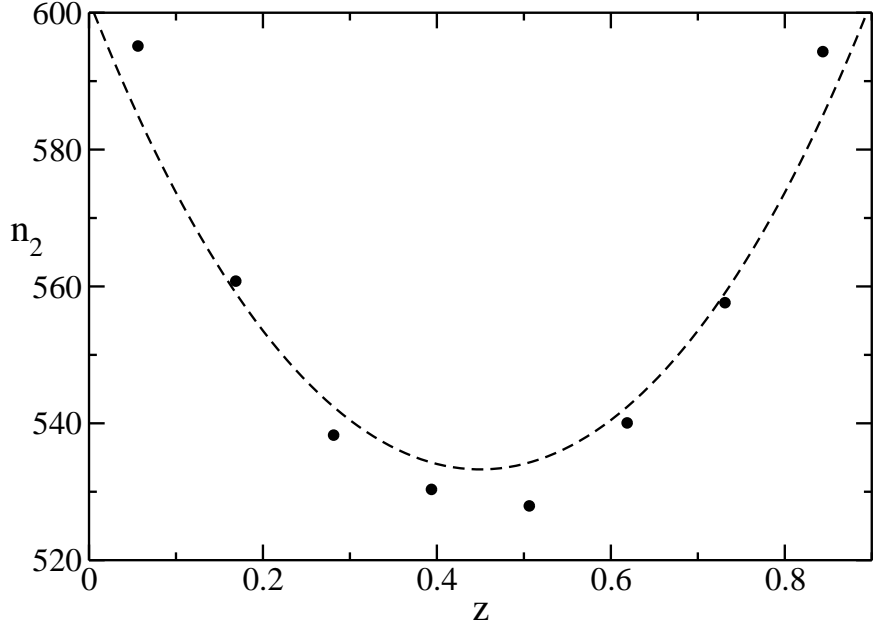}
  \caption{Density profile along the direction perpendicular to the plates in a confined quasi-two-dimensional system of hard spheres. The symbols are simulation results, while the dashed line is the theoretical prediction given by Eq. (\protect{\ref{17}}). The dimensionless density plotted is $n_{2}(z) \equiv n(z) \sigma A$  and the coordinate $z$ is measured in units of $\sigma$. The simulation data have been averaged on time, once the system is in the stationary state.}
  \label{fig2}
\end{figure}

The next aim of this paper is to show that an H theorem, similar to the Boltzmann result, but leading to an inhomogeneous steady state, is valid for the kinetic equation (\ref{1}), describing a confined gas. This result furnishes an example of the explicit proof of approach to  inhomogeneous equilibrium  of a strongly confined gas, hence bridging microscopic dynamics and macroscopic irreversibility. A non-equilibrium entropy is defined by
\begin{equation}
\label{3}
S(t) = -k_{B} H(t),
\end{equation}
where $k_{B}$ is the Boltzmann constant and $H(t)= H^{k}(t)+H^{c}(t)$, with the ``kinetic part'' $H^{k}$ having the Boltzmann form
\begin{equation}
\label{4} 
H^{ k}(t) \equiv \int d{\bm r} \int d{\bm v}\, f({\bm r},{\bm v},t) \left[ \ln f({\bm r},{\bm v},t) -1 \right]
\end{equation}
and the ``confinement contribution'' $H^{c}$  being
\begin{equation}
\label{5}
H^{c}(t) \equiv \frac{1}{2} \int d{\bm r} \int d{\bm r}_{1}\, n({\bm r},t) n(x,y,z_{1},t) \theta (\sigma -|{\bm r}_{1} -{\bm r}|).
\end{equation}  
In the last expression, $n({\bm r},t)$ is the local number density defined in the usual way
\begin{equation}
\label{6}
n({\bm r},t) \equiv \int d{\bm v}\, f({\bm r},{\bm v},t).
\end{equation}
The definition of $H^{c}(t)$ can be justified on the basis of a local equilibrium approximation for the general non-equilibrium $N$ particle distribution function of the system \cite{Re78a,Re78b}, keeping only contributions up to the second virial coefficient. This corresponds to taking the pair correlation function of the system equal to unity, but keeping the finite size of the particles in the description of the collision term. It is worth emphasizing that here the relevance of the size of the particles in the kinetic equation does not follow in principle from a density correction, but it appears as a consequence of the strong confinement of the system. For an homogeneous non-confined system, Eq. (\ref{5}) reduces to
\begin{equation}
\label{7} 
H^{c}_{hom} = Nn \frac{2\pi \sigma^{3}}{3},
\end{equation} 
that is just the second viral correction to the equilibrium entropy of a gas of hard spheres \cite{KyD71}.  Let us insist that Eq. (\ref{3}) is just a definition, valid for any arbitrary state defined by a solution of the kinetic equation and, therefore, it does not imply any approximation. Using a series of manipulations usual in hard-sphere dynamics \cite{McL89}, the time derivative of $H^{k}$  can be expressed as 
\begin{eqnarray}
\label{8}
\frac{\partial}{\partial t}  H^{k} &=& \frac{\sigma}{2} \int dx \int dy \int_{\sigma/2}^{h-\sigma/2} dz  \int_{\sigma/2}^{h-\sigma/2} dz_{1}  \int d{\bm v} \int d{\bm v}_{1}\, \int_{0}^{2\pi} d \varphi \, \nonumber \\
&&  
\times   \theta (- {\bm g} \cdot \widehat{\bm \sigma} ) |{\bm g} \cdot \widehat{\bm \sigma}|  f f_{1}  \ln \frac{f^{\prime} f^{\prime}_{1}}{f f_{1}},
 \end{eqnarray}
with $f \equiv  f({\bm r},{\bm v}, t)$, $f_{1} \equiv  f(x,y,z_{1},{\bm v}_{1}, t)$, $f^{\prime} \equiv  f({\bm r},{\bm v}^{\prime}, t)$, and $f^{\prime}_{1}  \equiv  f(x,y,z_{1},{\bm v}_{1}^{\prime},t)$. Upon deriving the above expression, it has been used that the term coming from free flow vanishes if one considers walls such that the flux of any property vanishes at them and that, for instance, periodic boundary conditions are considered in the $x$ and $y$ directions.  Now the inequality $x( \ln y-\ln x) \leq y-x$, valid for $x,y >0$ is employed. The equality sign only holds for $x=y$. After some straightforward algebra it is obtained
\begin{equation}
\label{9}
\frac{\partial}{\partial t}  H^{k} \leq I(t),
\end{equation}
\begin{equation}
\label{10}
I(t) \equiv 2 \pi \int dx \int dy \int_{\sigma/2}^{h-\sigma/2} dz \int_{\sigma/2}^{h-\sigma/2} dz_{1} n({\bm r},t) n(x,y,z_{1},t)(z_{1}-z) u_{z}(x,y,z_{1},t),
\end{equation}
where the local velocity field ${\bm u}({\bm r},t)$ defined by
\begin{equation}
\label{11}
n({\bm r},t){\bm u}({\bm r},t)  \equiv \int d{\bm v}\, {\bm v} f({\bm r},{\bm v},t),
\end{equation}
has been introduced. Equation (\ref{9}) has some similarities with the one obtained in \cite{Re78a,Re78b} in the context of the Enskog equation (for a non-confined fluid). The evolution equation for the confinement contribution to the entropy of the gas  is derived by means of the continuity equation
\begin{equation}
\label{12}
\frac{\partial}{\partial t} n({\bm r},t) = - \frac{\partial}{\partial {\bm r} } \left[ n({\bm r},t) {\bm u}({\bm r},t) \right],
\end{equation}
that follows directly from the kinetic equation (\ref{1}). The result is $ \partial H^{c}(t) / \partial t =-I(t)$, being $I(t)$ the same as given by Eq. (\ref{10}). Consequently, $\partial_{t}H(t) \leq 0$ or
\begin{equation}
\label{13} 
\frac{\partial S(t)}{\partial t} \geq 0 .
\end{equation}
Assuming as usual that the number density and the energy density are finite, it is easily shown that the expression of the entropy $S$, Eq. (\ref{3}), is bounded from above
\cite{vByE73} and, therefore, the distribution function tends towards a steady value in the long time limit. In the steady state $S$ must have a time-independent value and this only happens if the equality sign applies in Eq. (\ref{9}), i.e. if 
\begin{equation}
\label{14} 
\ln f +\ln f_{1} = \ln f^{\prime}+ \ln f^{\prime}_{1}.
\end{equation}
Proceeding similarly as it is done for the usual Boltzmann equation \cite{Ce88,McL89}, it is found that the only physically relevant steady  solution of the kinetic equation for the kind of boundary conditions we are considering is given by
\begin{equation}
\label{15}
f_{st}({\bf r},{\bm v})= n(z) \varphi_{MB}({\bm v}),
\end{equation}
where $\varphi_{MB}({\bm v})$ is the Maxwellian velocity distribution,
\begin{equation}
\label{16}
\varphi_{MB}({\bm v}) = \left( \frac{m}{2 \pi k_{B}T} \right)^{3/2}e^{-\frac{mv^{2}}{2k_{B}T}},
\end{equation}
with $T$ being the temperature, and $n(z)$ a density profile in the $z$ direction that is identified by substituting Eq. (\ref{15}) into the Boltzmann equation (\ref{1}), with the result
\begin{equation}
\label{17}
n(z) = \frac{N}{A b} \exp \left[ a \left( z - \frac{h}{2} \right)^{2} \right].
\end{equation}
Here $A$ is the area of each of the two parallel plates, $a\equiv \pi N /A$, 
\begin{equation}
\label{18}
b = \sqrt{\frac{\pi}{a}} \erfi \left[ \sqrt{a} \left( \frac{h}{2}- \frac{\sigma}{2} \right) \right],
\end{equation}
and $\erfi(x)=-i \erf (ix)$  denotes the imaginary error function of $x$. Note that $n(z)$ contains all the powers of the two-dimensional density through its dependence on $a$. To lowest order in the density, Eq. (\ref{17})  agrees with the low density limit of the expression derived in refs.
\cite{SyL96} and \cite{SyL97} starting from  the equilibrium BGY hierarchy for a system of hard spheres in presence of an external potential. Let us stress that the kinetic equation (\ref{1}) does not admit an homogeneous time independent solution. To check the accuracy of the above theoretical prediction, molecular dynamics simulations of a system of 500 hard spheres have been performed. The two confining plates are squares separated a distance  $h=1.9 \sigma$, and their area $A$ is such that $N \sigma^{2} /A
=0.19$. Figure \ref{fig2} displays the shape of the density profile along the $z$ direction. Although the density is not very low, the agreement between the theoretical prediction (solid line) and the simulation results (symbols) can be considered as quite satisfactory.

In the steady state, the entropy can be rewritten as
\begin{equation}
\label{19}
S_{st}= S_{st}^{id}+S_{st}^{ex},
\end{equation}
where $S_{st}^{id}$ is the ideal gas entropy
\begin{equation}
\label{20}
S^{id}_{st}= Nk_{B} \left( \frac{3}{2} \ln T- \ln n_{0}- \frac{3}{2} \ln \frac{m}{2 \pi k_{B}}+\frac{5}{2} \right),
\end{equation}
with $n_{0} \equiv N/A(h-\sigma)$, and the expression of the excess entropy reads
\begin{equation}
\label{21}
S_{st}^{ex}= -Nk_{B} \left( \ln \frac{h-\sigma}{b} + \frac{N \pi \sigma^{2}}{2A} \right).
\end{equation}
Again, the low density limit of this expression coincides with the low density limit of the result reported in ref. \cite{SyL97}. The force per unit of area exerted on the plates in the steady state is
\begin{equation}
\label{22}
p \equiv  \frac{T}{A} \left( \frac{\partial S_{st}}{\partial h} \right)_{N,T,A}= n(z=\sigma/2) k_{B} T,
\end{equation}
and the force per unit of length in the direction parallel to the walls has the form
\begin{equation}
\label{23}
\Sigma \equiv T \left( \frac{\partial S_{st}}{\partial A} \right)_{N,T,h}= k_{B} T \left[ \frac{3N}{2A} - \frac{h-\sigma}{2}\,  n(z=\sigma/2)+\frac{N^{2} \pi \sigma^{2}}{2 A^{2}} \right].
\end{equation}
The simplicity of Eq. (\ref{22}) reflects an ideal gas behavior in the transversal direction, consistently with the behavior found in the limit of extreme confinement $h \rightarrow 0$
\cite{FLyS12}. In the same limit, $\Sigma$ can be interpreted as the surface tension of a reference hard-disk system \cite{FLyS12}. To avoid misunderstandings, it is worth to note that the second term on the right hand side of Eq. (\ref{23}) reduces to $N/2A$ to lowest order in the density, then leading to the expected expression of $\Sigma$ as 
$k_{B}T N/A$.

In summary, the original Boltzmann ideas still remain valid in describing the kinetics and the approach to equilibrium in confined quasi-two-dimensional gases of hard spheres. 
The definition of entropy used is, perhaps, the simplest generalization of the original Boltzmann expression which leads to the right equilibrium entropy of a dilute confined gas, and it has been shown that it obeys an $H$ theorem. The kinetic equation reported here opens the way for the derivation of hydrodynamic-like macroscopic equations for the system, and to investigate the transition from three-dimensional to two-dimensional hydrodynamics. Moreover, and in relation with the experiments with macroscopic spheres mentioned at the beginning, the only modification needed to apply the kinetic equation  to the case of inelastic hard spheres refers to the collision rule, i.e. the expressions of the post-collisional velocities in terms of the pre-collisional ones and the scattering angle. In the last years, a general non-equilibrium statistical theory is being  developed for inelastic hard spheres \cite{BDyS97,Go03} that can be easily adapted to the present case.

This research was supported by the Ministerio de Econom\'{\i}a y Competitividad  (Spain) through Grant No. FIS2014-53808-P (partially financed by FEDER funds).

 \end{document}